\documentclass[11pt]{article}
\usepackage{amssymb,amsmath,epsfig}
\def\one{{{{\rm 1} \kern -.19em {\rm l}}}}
\def\C{{{{\rm {\mbox{\small l}}} \kern -.50em {\rm C}}}}
\def\R{{{{\rm l} \kern -.15em {\rm R}}}}
\def\N{{{{\rm l} \kern -.15em {\rm N}}}}
\def\E{{{{\rm l} \kern -.15em {\rm E}}}}
\def\P{{{{\rm l} \kern -.15em {\rm P}}}}
\def\Z{{{{\rm Z} \kern -.35em {\rm Z}}}}
\def\1{{{{\rm 1} \kern -.35em {\rm 1}}}}

\begin{document}
\begin{sloppypar}
\vspace*{0cm}
\begin{center}
{\setlength{\baselineskip}{1.0cm}{ {\Large{\bf
BOUND STATES OF THE TWO-DIMENSIONAL \\DIRAC EQUATION FOR AN ENERGY-DEPENDENT 
HYPERBOLIC SCARF POTENTIAL
\\}} }}
\vspace*{1.0cm}
{\large{\sc{Axel Schulze-Halberg}}}$^\dagger$ and {\large{\sc{Pinaki Roy}}}$^\ddagger$
\end{center}
\noindent \\

$\dagger$ Department of Mathematics and Actuarial Science and Department of Physics, Indiana University Northwest, 3400 Broadway,
Gary IN 46408, USA, e-mail: axgeschu@iun.edu, xbataxel@gmail.com \\ \\

$\ddagger$ Physics and Applied Mathematics Unit, Indian Statistical Institute, Kolkata 700108, India, 
e-mail: pinaki@isical.ac.in \\ \\

\vspace*{.5cm}
\begin{abstract}
\noindent
We study the two-dimensional massless Dirac equation for a potential that is allowed to depend on the energy 
and on one of the spatial variables. After determining a modified orthogonality relation and norm for such systems, 
we present an application involving an energy-dependent version of the hyperbolic Scarf potential. We 
construct closed-form bound-state solutions of the associated Dirac equation.

\end{abstract}


\section{Introduction}
The two-dimensional massless Dirac equation can be used to model electron transport phenomena in graphene, 
an atomically thin conducting material that consists of carbon atoms forming a honeycomb lattice 
structure \cite{novo}. This structure gives rise to many unusual properties of graphene, such as its 
very high electric conductivity, where both electrons and holes serve as charge carriers \cite{neto}. Low-energy 
states of electrons or holes in graphene can be modeled by the two-dimensional massless Dirac equation 
\cite{gonz}. In order to confine charge carriers, the Dirac equation must be coupled to a suitable scalar or 
vector potential. While for the vast majority of 
such external potentials the Dirac equation will not be solvable, there are some exceptions. Such 
exceptional cases include coupling to scalar potentials \cite{ghosh} \cite{ho} \cite{hartmann} \cite{xbat}, as well as to vector potentials 
\cite{check} \cite{jak} \cite{roy}. The purpose of this work is to extend the latter context to the case where the 
external potential depends on the energy. Quantum systems that feature energy-dependent potentials have been 
studied in the nonrelativistic case, for a comprehensive theoretical review and introductory examples 
the reader may refer to \cite{formanek} and \cite{sazdjian}. Energy-dependent potentials appear in a variety of 
applications, such as hydrodynamics \cite{hydro1}, confined quantum systems \cite{lombard} \cite{yekken2}, or 
multi-nucleon systems \cite{miya}. Theoretical applications include the generation of 
nonrelativistic models with energy-dependent potentials by means of the 
supersymmetry formalism \cite{yekken} and through point transformations \cite{jesus} \cite{xbatpct}. 
In the nonrelativistic context, the presence of energy-dependent potentials require a modification of the 
underlying quantum theory, principally affecting orthogonality relation and 
norm \cite{formanek}. In this note we show that a similar type of modification is also necessary in the 
relativistic context if the potential is energy-dependent. There is very few literature on the topic, particular 
systems were studied for example in \cite{hassan} \cite{ikot}, \cite{ikot2}. We consider here 
a particular case of the two-dimensional, massless 
Dirac equation for an external scalar potential that we assume to depend on the energy and on a single spatial variable. 
In section 2 we derive a modified orthogonality relation and norm for systems governed by such Dirac equations. 
Afterwards, we introduce an energy-dependent version of the hyperbolic Scarf potential. The conventional, 
energy-independent version of this potential was shown to support closed-form zero-energy states \cite{ghosh}. 
In section 3 we construct bound-state solutions of our Dirac equation for the energy-dependent hyperbolic 
Scarf potential and give several examples.

\section{The relativistic model}
We start out by introducing the two-dimensional, massless Dirac equation for an energy-dependent potential. 
The time-dependent version of this equation that we consider here features a potential that depends only on one 
of the spatial coordinates. It can be written in atomic units as follows
\begin{eqnarray}
\left[-i~\alpha \cdot \nabla+ V \hspace{-.1cm}\left(x,i \frac{\partial}{\partial t} \right) \right] \hat{\Psi}(x,y,t) &=& 
i~\frac{\partial \hat{\Psi}(x,y,t)}{\partial t},~~~~~(x,y,t) \in \mathbb{R}^2 \times (0,\infty), \label{tdirac} 
\end{eqnarray}
where $\alpha=(\sigma_1,\sigma_2)$ has the Pauli spin matrices $\sigma_1,\sigma_2$ as components and 
the potential $V$ is a continuous function of two variables. We can 
generate energy-dependence in the potential upon setting
\begin{eqnarray}
\hat\Psi(x,y,t) &=& \exp\left(-i~E~t \right) \Psi(x,y), \label{ts}
\end{eqnarray}
introducing the wave number $k_y$ that describes free motion in the $y$-direction, and the 
real-valued constant $E$ that will represent the stationary energy of our system. 
Insertion into (\ref{tdirac}) gives a stationary Dirac equation of the form
\begin{eqnarray}
\left\{-i~\alpha \cdot \nabla+ \left[V(x,E)-E \right] \right\} \Psi(x,y) &=& 0,~~~(x,y) \in \mathbb{R}^2. \label{dirac} 
\end{eqnarray}
We observe that the potential now depends on the energy $E$. 
Note further that, following the usual convention, we suppress the dependence on the energy $E$ in the solution 
$\Psi$ and its components.

\subsection{Derivation of orthogonality relation and norm}
It was shown \cite{formanek} that in the nonrelativistic 
scenario the presence of an energy-dependent potential forces a modification of the orthogonality relation and the 
norm, defined in the usual $L^2$-sense. For this reason, a similar modification is in order to accomodate systems 
governed by the Dirac equation (\ref{dirac}). Our starting point for the construction of a modified orthogonality relation 
and norm is the continuity equation
\begin{eqnarray}
\frac{\partial P(x,y,t)}{\partial t} &=& - \nabla J(x,y,t), \label{cont}
\end{eqnarray}
where $P$ and $J$ denote the relativistic probability density and probability current, respectively, that depend on 
the two spatial coordinates and on the time $t$. Let us now assume that $E_n$ and $E_m$ are two nonequal 
energies, for which the stationary Dirac equation (\ref{dirac}) admits solutions $\Psi_n$ and $\Psi_m$, respectively. 
The associated solutions of the time-dependent Dirac equation (\ref{tdirac}) can be found through (\ref{ts}). 
In addition, since the potential in our Dirac equation depends only on the $x$-coordinate, we can separate 
the $y$-coordinate off. More precisely, we set for $j=m$ and $j=n$
\begin{eqnarray}
\hat{\Psi}_j(x,y,t) ~=~ \exp\left(-i ~E_j~ t+i~k_y~y\right) \Psi_j(x), \label{hatpsi}
\end{eqnarray}
Keeping this relation in mind, we will now replace our standard continuity equation (\ref{cont}) by a modified version that 
satisfies the requirements imposed by an energy-dependent potential. Let us first define the probability 
density $P$ and the probability current $J$. These objects take the same form as in the conventional case, where the 
potential does not depend on the energy. We have
\begin{eqnarray}
P(x,y,t) &=& \hat{\Psi}_m^\dagger(x,y,t)~\hat{\Psi}_n(x,y,t) \label{prob} \\[1ex]
J(x,y,t) &=& \hat{\Psi}_m^\dagger(x,y,t)~\alpha~\hat{\Psi}_n(x,y,t). \label{curr}
\end{eqnarray}
Recall that the time-dependency of the expressions on the right sides is governed by (\ref{hatpsi}). We will now show that 
(\ref{prob}) and (\ref{curr}) satisfy the following modified continuity equation 
\begin{eqnarray}
\frac{\partial P(x,y,t)}{\partial t} +i\left[V(x,E_n)-V(x,E_m)\right] 
\hat{\Psi}_m^\dagger(x,y,t)~\hat{\Psi}_n(x,y,t)  &=& -\nabla J(x,y,t), \label{contmod}
\end{eqnarray}
where the symbol $\dagger$ denotes the hermitian adjoint. We observe that in contrast to the standard 
continuity equation (\ref{cont}), the modified version (\ref{contmod}) contains an additional term. The presence of 
this term is similar to the nonrelativistic scenario that was discussed in \cite{formanek} \cite{sazdjian}. 
Let us briefly show that (\ref{prob}) and (\ref{curr}) indeed satisfy the modified continuity equation 
(\ref{contmod}). To this end, we substitute (\ref{prob}) into the left side of the latter equation. For the sake of 
legibility we first evaluate the derivative with respect to the time variable. Taking into account the Dirac equation 
(\ref{dirac}) and using standard properties of the $\alpha$-matrix, this gives
\begin{eqnarray}
\frac{\partial P(x,y,t)}{\partial t} &=& \frac{\partial}{\partial t} 
\left[ \hat{\Psi}_m^\dagger(x,y,t)~\hat{\Psi}_n(x,y,t) \right] \nonumber \\[1ex]
&=& \frac{\partial \hat{\Psi}_m^\dagger(x,y,t)}{\partial t}~\hat{\Psi}_n(x,y,t)+
\hat{\Psi}_m^\dagger(x,y,t) ~\frac{\partial \hat{\Psi}_n(x,y,t)}{\partial t} \nonumber \\[1ex]
&=& \left[-\alpha \cdot \nabla \hat{\Psi}_m^\dagger(x,y,t)+i~V(x,E_m)~\hat{\Psi}_m^\dagger(x,y,t) \right]
\hat{\Psi}_n(x,y,t) + \nonumber \\[1ex]
&+& \hat{\Psi}_m^\dagger(x,y,t) 
\left[-\alpha \cdot \nabla \hat{\Psi}_n(x,y,t)-i~V(x,E_n)~\hat{\Psi}_n(x,y,t) \right] \nonumber \\[1ex]
& & \hspace{-3cm}=~-\left[\alpha \cdot \nabla \hat{\Psi}_m^\dagger(x,y,t) \right] \hat{\Psi}_n(x,y,t)
-\left[\alpha \cdot \nabla \hat{\Psi}_n(x,y,t) \right] \hat{\Psi}_m^\dagger(x,y,t) + i\left[V(x,E_m)-V(x,E_n) \right] \nonumber \\[1ex]
& & \hspace{-3cm}=~-\nabla \left[\hat{\Psi}_m^\dagger(x,y,t)~\alpha~\hat{\Psi}_n(x,y,t) \right]
 + i\left[V(x,E_m)-V(x,E_n) \right] \nonumber \\[1ex]
& & \hspace{-3cm}=~-\nabla J(x,y,t) + i\left[V(x,E_m)-V(x,E_n) \right]. \label{contmodcheck}
\end{eqnarray}
If we replace the derivative with respect to the time variable on the left side of (\ref{contmod}) by expression 
(\ref{contmodcheck}), we see that our modified continuity equation is satisfied. We are now able to construct an orthogonality relation for the solutions $\Psi_m$ and 
$\Psi_n$ of the stationary Dirac equation (\ref{dirac}). To this end, we integrate our continuity equation (\ref{contmod}) with 
respect to the time variable. Recall that the time-dependence of our spinors lies entirely in an exponential function 
as shown in (\ref{hatpsi}). Taking into account the latter definition in combination with (\ref{prob}), the left side of 
(\ref{contmod}) can be integrated as follows
\begin{eqnarray}
\int\limits^t \left\{ \left[\frac{\partial}{\partial s}~ \hat{\Psi}_m^\dagger(x,y,s)~\hat{\Psi}_n(x,y,s) \right]+i\left[V(x,E_n)-V(x,E_m)\right] 
\hat{\Psi}_m^\dagger(x,y,s)~\hat{\Psi}_n(x,y,s) \right\} ds ~=\nonumber \\[1ex]
& & \hspace{-9.7cm} =~ \left[1-\frac{V(x,E_m)-V(x,E_n)}{E_m-E_n}\right] 
\hat{\Psi}_m^\dagger(x,y,t)~\hat{\Psi}_n(x,y,t). \label{intprob}
\end{eqnarray}
At this point it is convenient to make use of the relation (\ref{hatpsi}) by substituting it into (\ref{intprob}). 
Similar to the nonrelativistic case \cite{formanek} \cite{sazdjian}, this leads to the sought orthogonality relation 
\begin{eqnarray}
\int\limits_{\mathbb{R}} \left[1-\frac{V(x,E_m)-V(x,E_n)}{E_m-E_n}\right] \Psi_m^\dagger(x)~\Psi_n(x)~dx &=& 
C~\delta_{mn}, \label{ortho}
\end{eqnarray}
where $C$ is a constant. From the orthogonality relation (\ref{ortho}) we can now 
construct the modified norm $N$ by taking the limit $m \rightarrow n$, resulting in
\begin{eqnarray}
N \hspace{-.1cm} \left(\Psi_n \right) &=& \int\limits_{\mathbb{R}} \left[1-\frac{\partial V(x,E_n)}{\partial E_n}\right] 
\Psi_n^\dagger(x)~\Psi_n(x)~dx. 
\label{norm}
\end{eqnarray}
For a solution $\Psi_n$ of the Dirac equation (\ref{dirac}) to represent a bound state, two conditions must be 
fulfilled: the norm integral (\ref{normx}) must exist and at the same time its integrand must be a nonnegative 
function. Since the sign of the integral is entirely determined by the factor involving the potential's derivative, the 
condition 
\begin{eqnarray}
1-\frac{\partial V(x,E_n)}{\partial E_n} &\geq& 0, \label{sign}
\end{eqnarray}
must be satisfied for all real numbers $x$ and energies $E_n$ associated with the system governed by the Dirac equation (\ref{dirac}). 
Finally let us 
note that (\ref{norm}) does not constitute a norm in the mathematical sense because it can become 
negative due to the term containing the energy derivative of the potential.

\subsection{Decoupling the Dirac system}
Before we can consider applications involving specific energy-dependent potentials, we must solve the Dirac 
equation (\ref{tdirac}). Since this equation has two components, it can be written as a system of two equations that 
must be decoupled. To this end, we represent the solution spinor $\hat{\Psi}$ in the form (\ref{hatpsi}) and split it up 
into its two components. We set
\begin{eqnarray}
\hat{\Psi}(x,y,t) &=&  \frac{1}{2}~\exp\left(-i~E_n~t+i~k_y~y \right)
\left(
\begin{array}{ll}
\psi_+(x) \\[1ex]
\psi_-(x)
\end{array}
\right) \nonumber \\[1ex]
&=&
\frac{1}{2}~\exp\left(-i~E_n~t+i~k_y~y \right)
\left(
\begin{array}{ll}
\psi_{1}(x)+\psi_{2}(x) \\[1ex]
\psi_{1}(x)-\psi_{2}(x)
\end{array}
\right). \label{psinew}
\end{eqnarray}
Note that the factor $1/2$ was introduced merely to facilitate calculations. 
Upon substitution of (\ref{psinew}) into (\ref{tdirac}), we obtain the following system of equations \cite{ghosh} for the spinor components $\psi_{1}$ and 
$\psi_{2}$. 
\begin{eqnarray}
\psi_{1}''(x)+\Big\{\left[V(x,E)-E \right]^2+i~\frac{\partial V(x,E)}{\partial x}-k_y^2\Big\}~ \psi_{1}(x) ~=~0 \label{eq1} 
\end{eqnarray}
\vspace{-.5cm}
\begin{eqnarray}
\hspace{-.85cm} \psi_{2}(x) ~=~ \frac{1}{k_y} ~\Big\{\psi_{1}'(x)+i \left[V(x,E)-E \right] \psi_{1}(x)\Big\},  \label{psi2}
\end{eqnarray}
where we assume without restriction that $k_y \neq 0$. Once the first solution component $\psi_{1}$ has been found from the Schr\"odinger-type equation (\ref{eq1}), 
the remaining counterpart $\psi_{2}$ is generated by means of (\ref{psi2}). These two functions are then substituted into 
(\ref{psinew}) in order to obtain the solution spinor of the stationary Dirac equation. Before we apply the results of this 
section to a specific model, we rewrite the orthogonality relation (\ref{ortho}) and norm (\ref{norm}) in terms of 
the solutions to the system (\ref{eq1}), (\ref{psi2}). To this end, we introduce two pairs of functions 
$\psi_{1,m}$, $\psi_{2,m}$ and $\psi_{1,n}$, $\psi_{2,n}$ that are solutions to the latter system for $E=E_m$ and 
$E=E_n$, respectively. Upon substituting relation (\ref{psinew}) into our orthogonality relation (\ref{ortho}) and 
norm (\ref{norm}), we obtain the results
\begin{eqnarray}
& & \int\limits_{\mathbb{R}} \left[1-\frac{V(x,E_m)-V(x,E_n)}{E_m-E_n}\right] \Bigg\{\left[\psi^\ast_{1,m}(x)+\psi^\ast_{2,m}(x) \right]
\left[\psi_{1,n}(x)+\psi_{2,n}(x) \right]+ \nonumber \\[0ex]
& & \hspace{4.55cm}+\left[\psi^\ast_{1,m}(x)-\psi^\ast_{2,m}(x) \right]
\left[\psi_{1,n}(x)-\psi_{2,n}(x) \right]\Bigg\}~ dx ~=~ 
C~\delta_{mn} \nonumber \\[1ex]
& & N \hspace{-.1cm} \left(\psi_n \right) ~=~ \int\limits_{\mathbb{R}} \left[1-\frac{\partial V(x,E_n)}{\partial E_n}\right] 
\Big[|\psi_{1,n}(x)+\psi_{2,n}(x)|^2+
|\psi_{1,n}(x)-\psi_{2,n}(x)|^2\Big] dx. \label{normx}
\end{eqnarray}
Observe that the dependence on the spatial variable $y$ is gone because the corresponding exponential term 
from (\ref{hatpsi}) cancels out, leaving a single integral. Note further that we left some irrelevant overall constant 
factors out.

\section{Application: hyperbolic Scarf potential}
We will now introduce a particular energy-dependent potential, for which our stationary Dirac equation (\ref{dirac}) 
admits bound-state solutions that can be given in closed form. The potential we will focus on reads
\begin{eqnarray}
V(x,E) &=& -\lambda(E)~\mbox{sech}(x)+\mu(E)~\tanh(x)+E, \label{v}
\end{eqnarray}
where $\lambda \neq 0$ and $\mu$ are real-valued functions of the energy parameter $E$. 
We see that the function (\ref{v}) is an energy-dependent generalization of the hyperbolic Scarf potential. 
It is known \cite{ghosh} that the conventional, energy-indepedent version of our potential represents a 
well for electrons if $\lambda>0$ and a well for holes if $\lambda<0$. In what follows we will show that this 
interpretation can be maintained if the potential is energy-dependent and of the form (\ref{v}), provided 
certain constraints are met.  Observe that the last term on the right side of (\ref{v}) will cancel with the same term 
in our Dirac equation (\ref{dirac}). As such, solutions of the latter equation are formally equivalent to zero-energy 
solutions for the scenario of an energy-independent potential. Let us further remark that the potential (\ref{v}) has a formal similarity with the 
potential discussed in \cite{hartmann2}, as far as the shape of its graph is concerned. However, the latter 
reference considers the conventional, energy-independent context.

\subsection{General solution of the governing equation}
Our starting point is the observation that the Dirac equation (\ref{dirac}) for our 
potential (\ref{v}) is exactly-solvable. The general solution provided 
in \cite{ghosh} persists under the generalization regarding the energy-dependent parameters. Since the explicit form of 
the solution spinor (\ref{psinew}) is too long to be shown here, we focus on the function $\psi_1$ that is determined by the 
Schr\"odinger-type equation (\ref{eq1}). This equation reads after incorporation of (\ref{v})
\begin{eqnarray}
\psi_1''(x)+\Bigg\{-k_y^2+\mu(E)^2+\mbox{sech}^2(x)\Big[\lambda(E)^2+i~\mu(E)-\mu(E)^2 \Big]
+\mbox{sech}(x)~\tanh(x) \Big[i~\lambda(E)-  & &  \nonumber \\[1ex]
& & \hspace{-6.0cm} -~2~\lambda(E)~\mu(E) \Big] \Bigg\} ~\psi_1(x) ~=~ 0. \label{eqpre}
\end{eqnarray}
The general solution of this equation for $\psi_{1}$ is given by
\begin{eqnarray}
\psi_{1}(x)  \hspace{-.1cm} &=& \hspace{-.1cm}c_1~\left[1-i~\sinh(x) \right]^{\frac{c}{2}-\frac{1}{4}} 
~\left[1+i~\sinh(x) \right]^{\frac{a}{2}+\frac{b}{2}-\frac{c}{2}+\frac{1}{4}}~
{}_2F_1\left[
a,b,c,\frac{1}{2}-\frac{i}{2}~\sinh(x)\right]+ \nonumber \\[1ex]
& & \hspace{-1.5cm}+~c_2~
\left[1-i~\sinh(x) \right]^{\frac{3}{4}-\frac{c}{2}} ~
\left[1+i~\sinh(x) \right]^{\frac{1}{4}-\frac{a}{2}-\frac{b}{2}+\frac{c}{2}}~
{}_2F_1\left[
1-a,1-b,2-c,\frac{1}{2}-\frac{i}{2}~\sinh(x)\right] \hspace{-.1cm}. \nonumber \\[1ex] \label{psi1x}
\end{eqnarray}
Here, $c_1, c_2$ are arbitrary constants and ${}_2F_1$ stands for the hypergeometric function \cite{abram}. 
Furthermore, the following abbreviations are in use
\begin{eqnarray}
a&=&\frac{1}{2}-\frac{1}{4}~\sqrt{\left[-1+2~\lambda(E)-2~i~\mu(E)\right]^2}
-\frac{1}{4}~\sqrt{\left[1+2~\lambda(E)+2~i~\mu(E)\right]^2}+\sqrt{k_y^2-\mu(E)^2}
\nonumber \\[1ex]
b&=&\frac{1}{2}-\frac{1}{4}~\sqrt{\left[-1+2~\lambda(E)-2~i~\mu(E)\right]^2}
-\frac{1}{4}~\sqrt{\left[1+2~\lambda(E)+2~i~\mu(E)\right]^2}-\sqrt{k_y^2-\mu(E)^2}
\nonumber \\[1ex]
c&=&1-\frac{1}{2}~\sqrt{\left[1+2~\lambda(E)+2~i~\mu(E)\right]^2}. 
\label{abc}
\end{eqnarray}
These expressions can be simplified further once the sign of the radicands is known. We will discuss this in detail 
further below. Observe that in (\ref{abc}) we did not include an argument to indicate the dependency of $a,b$ and $c$ on the 
energy $E$. We note that the function $\psi_2$ in (\ref{psinew}) can now be obtained from 
(\ref{psi1x}) through the relation (\ref{psi2}), determining the general solution of our Dirac equation (\ref{dirac}) 
for the potential (\ref{v}).

\subsection{Construction of bound states}
We will now 
impose additional conditions on (\ref{psi1x}) in order to extract bound-state solutions and their corresponding 
energies. For such solutions, the norm integral (\ref{normx}) must exist and the 
sign condition (\ref{sign}) must be fulfilled. We will investigate these two aspects separately.

\paragraph{Existence of the norm integral.} Let us first ensure that the norm (\ref{normx}) exists in the 
present case. To this end, we observe that our solution (\ref{psi1x}) becomes in general unbounded at the infinities due to 
the hypergeometric functions it contains. We modify the latter solution by setting $c_1=1$ and $c_2=0$, 
removing its second term on the right side. Next, we recall that the hypergeometric function degenerates to 
a polynomial if its first argument equals a nonpositive integer. Taking into account that this argument is given by $a$ and 
defined in (\ref{abc}), we obtain the constraint
\begin{eqnarray}
\frac{1}{2}-\frac{1}{4}~\sqrt{\left[1+2~\lambda(E)+2~i~\mu(E)\right]^2}
-\frac{1}{4}~\sqrt{\left[-1+2~\lambda(E)-2~i~\mu(E)\right]^2}+\sqrt{k_y^2-\mu(E)^2}
 = -n, \nonumber \\ \label{eigenpre}
\end{eqnarray}
for a nonnegative integer $n$. Since the first two complex roots on the left side can take two values each, 
we can simplify (\ref{eigenpre}) by distinguishing four possible cases, depending on the sign of the two roots. 
These cases are
\begin{eqnarray}
\frac{1}{2}-\lambda(E)+\sqrt{k_y^2-\mu(E)^2}
 = -n \label{eigen1} \\[1ex]
\frac{1}{2}+\lambda(E)+\sqrt{k_y^2-\mu(E)^2}
 = -n \label{eigen2} \\[1ex]
1+i~\mu(E)+\sqrt{k_y^2-\mu(E)^2}
 = -n \label{eigen3} \\[1ex]
-i~\mu(E)+\sqrt{k_y^2-\mu(E)^2}
 = -n. \label{eigen4}
\end{eqnarray}
Next, let us show that the last two cases can be discarded. To this end, we first assume that the root in (\ref{eigen3}) and 
(\ref{eigen4}) is real-valued. This implies $\mu=0$, such that the energy $E$ completely disappears from the condition. 
As a consequence, no stationary energies can be determined. If we assume that the root in (\ref{eigen3}) 
is imaginary, then (\ref{eigen3}) results in $n=-1$, which is not a valid assignment due to the restriction that 
$n$ must be a nonnegative integer. Finally, if the root in (\ref{eigen4}) takes imaginary values, we obtain $n=0$ and 
$|\mu(E)|=k_y/\sqrt{2}$. While this is in principle acceptable, we will see below that bound states of our system 
can only be constructed if $\mu$ is a constant. As such, the energy $E$ will once more disappear from our 
condition (\ref{eigen4}). For these reasons we only retain the conditions (\ref{eigen1}) and (\ref{eigen2}) that 
were obtained by assuming that the first two complex roots on the left side of (\ref{eigenpre}) take the same sign. 
Now, our condition (\ref{eigenpre}) can be rewritten using (\ref{eigen1}) and 
(\ref{eigen2}) as follows
\begin{eqnarray}
\frac{1}{2}-\epsilon~ \lambda(E_n)+\sqrt{k_y^2-\mu(E_n)^2}
 ~=~ -n \qquad\mbox{and} \qquad |k_y|~\geq~|\mu(E_n)|, \label{eigen}
\end{eqnarray}
where $n$ is a nonnegative integer and the new parameter $\epsilon$ can take either the value positive one or negative one. 
Since our potential coefficients $\lambda$ and $\mu$ depend on the stationary energy, we cannot determine an 
explicit formula for those energies from (\ref{eigen}), unless more information about the coefficients is known. 
Before we continue, a general remark on the role of the parameter $k_y$ is in order. We observe that a 
condition is placed on $k_y$ in order for (\ref{eigen}) to deliver real-valued energies. While this condition on $k_y$ 
is relatively mild, there are cases where stronger constraints are imposed. A typical example of such a case is the 
work \cite{ghosh}, where bound-state solutions of the Dirac equations are sought at zero energy. It is found that 
bound states can be constructed provided $k_y$ is constrained to attain certain values. This type of constraint does not 
appear in the present case because we do not set the energy to a fixed value. Let us for now assume that (\ref{eigen}) is satisfied. The stationary energies defined in the latter condition belong to 
bound-state type solutions of (\ref{eqpre}), given by the functions
\begin{eqnarray}
\psi_{1,n}(x) &=& 
\left[1-i~\sinh(x)\right]^{\frac{1}{4}-\frac{\epsilon}{4} \left[1+2 \lambda(E)+2 i \mu(E)\right]}
\left[1+i~\sinh(x)\right]^{\frac{1}{4}- \frac{\epsilon}{4} \left[-1+2 \lambda(E)-2 i \mu(E)\right]} \times \nonumber \\[1ex]
&& \hspace{-1.8cm} \times~ 
{}_2F_1\left[-n,-n-2~\sqrt{k_y^2-\mu(E)^2},1-\frac{\epsilon}{2}~\sqrt{\left[1+2~\lambda(E)+2~i~\mu(E)\right]^2},
\frac{1}{2}-\frac{i}{2} \sinh(x)
\right], \label{psihyp}
\end{eqnarray}
where irrelevant overall factors were omitted. Since the first argument of the hypergeometric function in (\ref{psihyp}) is 
a nonpositive integer, we can express it as follows
\begin{eqnarray}
\psi_{1,n}(x) &=& 
\left[1-i~\sinh(x)\right]^{\frac{1}{4}-\frac{\epsilon}{4}\left[1+2 \lambda(E)+2 i \mu(E)\right]}
\left[1+i~\sinh(x)\right]^{\frac{1}{4}-\frac{\epsilon}{4}\left[-1+2 \lambda(E)-2 i \mu(E)\right]} \times \nonumber \\[1ex]
&& \hspace{-1.5cm} \times~ 
P_n^{\left(-\frac{\epsilon}{2}\sqrt{\left[1+2~\lambda(E)+2~i~\mu(E)\right]^2},-\frac{\epsilon}{2}\sqrt{\left[1-2~\lambda(E)+2~i~\mu(E)\right]^2}
\right)}\left[i \sinh(x) \right]. \label{psi1bound}
\end{eqnarray}
Here, the symbol $P_n$ stands for a Jacobi polynomial of degree $n$. Before we continue, let us point out that the 
functions (\ref{psi1bound}) do not lead to bound states of our Dirac equation (\ref{dirac}) unless the parameters 
satisfy certain conditions. In particular, existence and positiveness of our norm (\ref{normx}) is not guaranteed in 
general. In order to find out more about this, let us now analyze the asymptotic behavior 
of the solution (\ref{psi1bound}) at the infinities. For the sake of simplicity we first assume that $n=0$, such that 
the Jacobi polynomial becomes equal to one and only two factors on the right side of (\ref{psi1bound}) remain. 
The asymptotics of these factors for large values of $|x|$ is determined by the real parts of their exponents. 
In particular, at least one of the exponents must have negative real part. In case one of the exponents has positive real part, 
it must be less than the absolute value of its counterpart. For $n=0$ and $\epsilon=1$, (\ref{psi1bound}) simplifies to
\begin{eqnarray}
\psi_1(x) &=& 
\left[1-i~\sinh(x)\right]^{-\frac{\lambda(E)}{2}-\frac{i \mu(E)}{2}}
\left[1+i~\sinh(x)\right]^{\frac{1-\lambda(E)}{2}+\frac{i \mu(E)}{2}}. \label{psic1}
\end{eqnarray}
We see that the exponents satisfy our requirements if the condition $\lambda(E) > \frac{1}{2}$ is satisfied. 
Let us now evaluate (\ref{psi1bound}) for $n=0$ and $\epsilon=-1$. This gives
\begin{eqnarray}
\psi_1(x) &=& 
\left[1-i~\sinh(x)\right]^{\frac{1+\lambda(E)}{2}+\frac{i \mu(E)}{2}}
\left[1+i~\sinh(x)\right]^{\frac{\lambda(E)}{2}-\frac{i \mu(E)}{2}}. \label{psic2}
\end{eqnarray}
This time we see that our requirements on imply $\lambda(E) < -\frac{1}{2}$. 
If we now drop the assumption of vanishing $n$, we must consider the effect that the Jacobi polynomial in (\ref{psi1bound}) 
has on our conditions for $\lambda$. To this end, we observe that the Jacobi polynomial depends on the variable $x$ 
merely through the term $1-i \sinh(x)$. As such, the degree of the polynomial adds $n$ to the exponent of the first factor 
on the right side of (\ref{psic1}) and (\ref{psic2}), respectively. As a direct consequence, our constraint on $\lambda$ 
becomes for $\epsilon=1$ 
\begin{eqnarray}
\lambda(E) &>& n+\frac{1}{2}. \label{l1}
\end{eqnarray}
This generalizes a condition found in \cite{ghosh} to the energy-dependent potential (\ref{v}). If (\ref{l1}) is satisfied, 
bound-state solutions of the Dirac equation (\ref{dirac}) describe the behavior of electrons. Next, if we choose 
$\epsilon=-1$, we arrive at the condition
\begin{eqnarray}
\lambda(E) &<& -n-\frac{1}{2}, \label{l2}
\end{eqnarray}
Similar to (\ref{l1}), this is a generalization of a constraint valid for the energy-independent version of (\ref{v}). If 
(\ref{l2}) holds, then bound-state solutions of the Dirac equation (\ref{dirac}) model the behavior of holes \cite{ghosh}. 
Hence, if either condition (\ref{l1}) or (\ref{l2}) are satisfied, then the corresponding solution in 
(\ref{psi1bound}) vanishes at both infinities. Furthermore, the derivative of (\ref{psi1bound}) with respect to $x$ 
shows the same behavior, implying that the remaining component (\ref{psi2}) forming our 
solution spinor (\ref{psinew}) vanishes at both infinities. For the sake of brevity we omit to show this 
rigorously, as it would require a similar series of considerations as done above for the function (\ref{psi1bound}). 
It now follows that the density $|\psi_+|^2+|\psi_-|^2$ also vanishes at the infinities, establishing existence of the 
norm integral in (\ref{normx}).

\paragraph{Sign of the norm integral.}
It now remains to study the sign of the norm (\ref{normx}) in order to ensure that condition (\ref{sign}) is fulfilled. 
The expression on the left side of this condition reads after substitution of our potential (\ref{v})
\begin{eqnarray}
1-\frac{\partial V(x,E)}{\partial E} &=& \lambda'(E)~\mbox{sech}(x)-\mu'(E)~\tanh(x). \label{ve}
\end{eqnarray}
Note that for a fixed value of $E$, this expression stays bounded on the whole real line, such that it cannot affect 
existence of the integral (\ref{normx}). In order to satisfy condition (\ref{sign}), the right side of (\ref{ve}) must be 
nonnegative for all $x$ and all admissible values of $E$. This 
is only possible if the coefficient of the hyperbolic secant is positive and if the hyperbolic tangent 
term is not present. In other words, we must impose the simultaneous conditions 
\begin{eqnarray}
\lambda'(E)~>~0 ~~~~\mbox{and}~~~~ \mu(E)~=~\mbox{constant}, \label{normex}
\end{eqnarray}
for all values of the energy $E$. In summary, if one of the conditions (\ref{l1}) or (\ref{l2}) is met and if in addition 
the sign condition (\ref{normex}) is fulfilled, then the functions (\ref{psi1bound}) generate bound-state solutions of 
the Dirac equation (\ref{dirac}) with energy-dependent potential (\ref{v}) by means of (\ref{psi2}) and (\ref{psinew}).

\paragraph{The case \boldmath{$\mu=0$}.} Before we conclude this section, let us briefly comment on a 
particular case of our potential (\ref{v}) that arises if the parameter $\mu$ vanishes. The resulting potential, 
consisting of a single secant term, satisfies the sign condition (\ref{normex})  and as such 
allows for the construction of bound-state solutions to our Dirac equation (\ref{dirac}). The secant potential 
is of importance especially in applications of graphene, as it was shown to match the graphene 
top-gate structure \cite{hart} \cite{hartmann}. In the latter references, bound-state solutions of the Dirac equation 
for an energy-independent secant potential were obtained, within a quasi-exactly solvable setting and at 
zero energy, respectively. Let us add that zero-energy solutions of the Dirac equation have also been found for 
different types of potentials, see for example \cite{down}. In the present case of energy-dependence 
in the potential, bound-state solutions and 
their associated stationary energies can be obtained directly from (\ref{psi1bound}) and (\ref{eigen}), respectively, 
by setting $\mu=0$. We will comment on this case below when discussing examples. For small values of $\mu$, 
the hyperbolic Scarf potential (\ref{v}) is a deformation of the secant potential, which is an even function. Due to this 
property, the effective complex potential in the Schr\"odinger-type equation (\ref{eqpre}) features 
${\cal PT}$-symmetry \cite{bender}.

\subsection{Applications}
Even though we were able to construct the general form of bound-state solutions through (\ref{psi1bound}), 
we can only give an implicit equation (\ref{eigen}) for the associated stationary energies. This changes once 
more information is known about the parameter functions $\lambda$ and $\mu$. Therefore, we will now choose 
particular cases of those functions and apply the results of the previous sections. 
It will turn out that, depending on the parameter values, the resulting stationary energies form infinite sequences that can be unbounded or 
have an accumulation point.

\subsubsection{Linear energy-dependence}
\begin{figure}[h]
\begin{center}
\epsfig{file=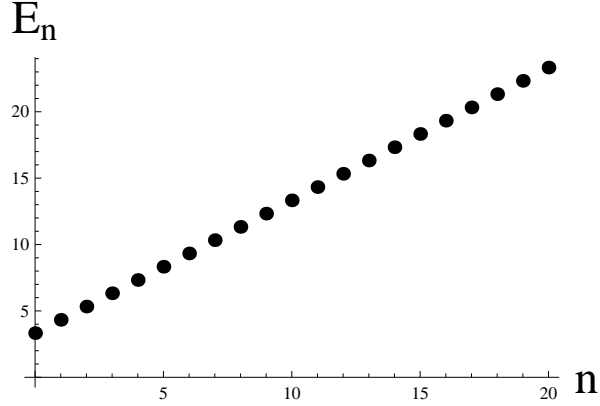,width=7.8cm}
\caption{The stationary energies (\ref{eigenx1}) for $0 \leq n \leq 20$. 
Parameter settings are $\alpha=1$, $k_y=2$ and $\beta=3/2$.}
\label{fig3a}
\end{center}
\end{figure}

\begin{figure}[h]
\begin{center}
\epsfig{file=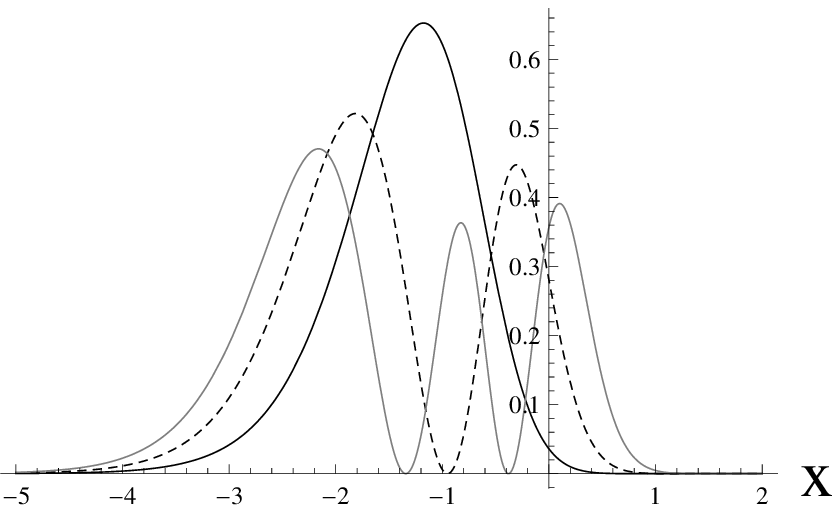,width=7.8cm}
\epsfig{file=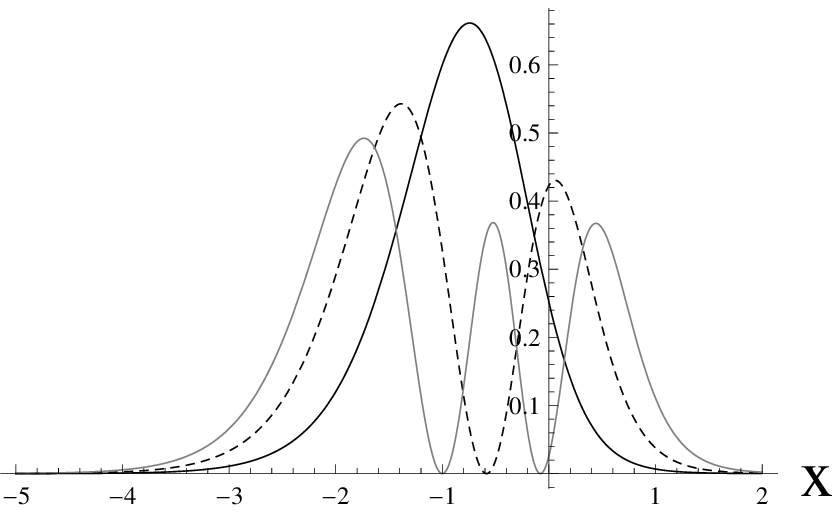,width=7.8cm}
\caption{The densities $|\psi_+|^2$ (left plot) and $|\psi_-|^2$ (right plot) for the 
values $n=0$ (black curve), $n=1$ (dashed curve), and $n=2$ (gray curve), respectively. 
Parameter settings are $\alpha=1$, $k_y=2$ and $\beta=3/2$.}
\label{fig1}
\end{center}
\end{figure}

In our first example let us employ the following settings
\begin{eqnarray}
\lambda(E) ~=~ \alpha~E \qquad \qquad \qquad \mu(E) ~=~ \beta, \label{set1}
\end{eqnarray}
where $\alpha>0$ and $\beta$ are real-valued constants. We observe that these settings are compatible with the 
requirement (\ref{normex}), a necessary condition for the construction of bound states. Furthermore, the constant 
$\beta$ is allowed to vanish, in which case (\ref{v}) turns into the hyperbolic secant potential. The remaining conditions for 
norm existence will be verified further below. Upon substitution of (\ref{set1}) into the potential (\ref{v}), we obtain
\begin{eqnarray}
V(x,E) &=& -\alpha~E~\mbox{sech}(x)+\beta~\tanh(x)+E. \label{v1}
\end{eqnarray}
This potential depends linearly on the energy in its first and third term. Next, let us determine the 
stationary energies of the system governed by the Dirac equation (\ref{dirac}) and the potential (\ref{v1}). 
These energies can be found from equation (\ref{eigen}), where we must first provide a value for $\epsilon$. This 
value depends on which of the two conditions (\ref{l1}), (\ref{l2}) we want to satisfy. We choose the first of these 
conditions, implying that $\epsilon=1$. Upon substitution of this value in combination with (\ref{set1}) into 
(\ref{eigen}), we obtain the following condition
\begin{eqnarray}
\frac{1}{2}- \alpha~E+\sqrt{k_y^2-\beta^2}
 ~=~ -n,~~~|k_y|~\geq~\beta. \nonumber
\end{eqnarray}
Solving for $E$ will give an explicit formula for our stationary energies. In order to indicate this, we 
amend $E$ by an index $n$, arriving at
\begin{eqnarray}
E_n &=& \frac{1}{\alpha} \left(n+\frac{1}{2}+\sqrt{k_y^2-\beta^2}\right), ~~~|k_y|~\geq~\beta. \label{eigenx1}
\end{eqnarray}
These values are positive and increase linearly with $n$. Figure \ref{fig3a} shows the lowest stationary energies 
for a particular setting of our parameters in (\ref{set1}).
Next, we must check for which values of the parameters $\alpha$, $\beta$ these energies comply with the 
existence condition (\ref{l1}) of the norm. Note that due to 
our parameter choice $\epsilon=1$, we do not consider the second condition (\ref{l2}). Taking into account the 
definition of $\lambda$ in (\ref{set1}), we substitute (\ref{eigenx1}) into (\ref{l1}). This gives 
$\lambda(E_n) > n + 1/2$, so that after simplification we obtain
\begin{eqnarray}
\sqrt{k_y^2-\beta^2} &>& 0. \label{true}
\end{eqnarray}
This constraint is satisfied since it coincides with our requirement in (\ref{eigenx1}).  Since condition (\ref{normex}) is already satisfied, 
it follows that the stationary energies (\ref{eigenx1}) are associated with solutions that are normalizable in the 
sense (\ref{normx}). 
There are infinitely many such bound-state solutions because the constraint (\ref{true}) is fulfilled for all values of $n$. 
Since the closed form of the bound-state solutions is very long, we restrict ourselves to showing only the 
function $\psi_1$. To this end, we insert the settings (\ref{set1}) and (\ref{eigenx1}) into (\ref{psi1bound}), arriving at the 
result
\begin{eqnarray}
\psi_{1,n}(x) &=& 
\left[1-i~\sinh(x)\right]^{\frac{1}{4}-\frac{1}{2}\left(1+n+\sqrt{k_y^2-\beta^2}+4 i \beta\right)}
\left[1+i~\sinh(x)\right]^{\frac{1}{4}-\frac{1}{2}\left(n+\sqrt{k_y^2-\beta^2}-4 i \beta\right)} \times \nonumber \\[1ex]
&& \hspace{-1.5cm} \times~ 
P_n^{\left(-\frac{1}{2}\sqrt{\left(2+2n+2\sqrt{k_y^2-\beta^2})+2 i \beta\right)^2},
-\frac{1}{2}\sqrt{\left(-2 n-2\sqrt{k_y^2-\beta^2}+2~i~\beta\right)^2}
\right)}\left[i \sinh(x) \right]. \label{psi1}
\end{eqnarray}
Note that we included the parameter $n$ as an index in order to indicate the bound-state character. In order to 
construct the solution to the Dirac equation (\ref{dirac}), we calculate $\psi_{2,n}$ by means of 
(\ref{psi2}). After that, we can calculate the norm of these bound-state solutions by means of (\ref{normx}). 
Since we know that the norm integral exists, it remains to  ensure that it gives a nonnegative result. To this end, 
we recall that the sign of the norm is determined by expression (\ref{ve}). In the present case, this expression is 
obtained by substituting (\ref{set1}) and evaluating the derivatives, giving 
\begin{eqnarray}
1-\frac{\partial V(x,E_n)}{\partial E_n} &=& \alpha~\mbox{sech}(x). \label{ve1}
\end{eqnarray}
Since both the constant $\alpha$ and the hyperbolic secant function are positive, the norm (\ref{normx}) of our bound-state 
solutions generated by (\ref{psi1}) will also be positive. Now that we have found the functions 
$\psi_{1,n}$ and $\psi_{2,n}$, we can determine the components $\psi_+$ and $\psi_-$ of the Dirac spinor (\ref{psinew}) through 
addition and subtraction, respectively. Figure \ref{fig1} shows these components for a particular parameter setting 
and the first values of $n$. Note that we normalized the functions shown in the figure, such that $|\psi_+|^2=|\psi_-|^2=1$.

\subsubsection{Inverse-power energy-dependence}
\begin{figure}[h]
\begin{center}
\epsfig{file=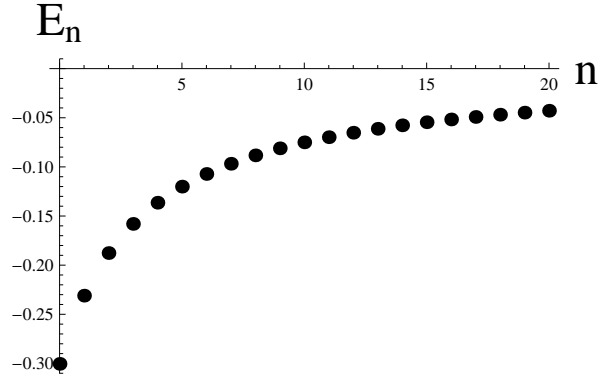,width=7.8cm}
\caption{The stationary energies (\ref{eigenx2}) for $0 \leq n \leq 20$. 
Parameter settings are $\alpha=\beta=1$, $k_y=3$ and.}
\label{fig3b}
\end{center}
\end{figure}

\begin{figure}[h]
\begin{center}
\epsfig{file=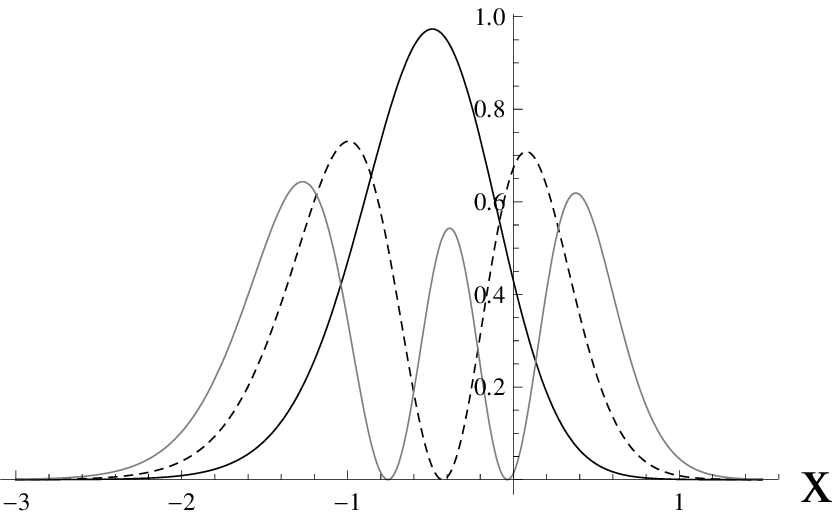,width=7.8cm}
\epsfig{file=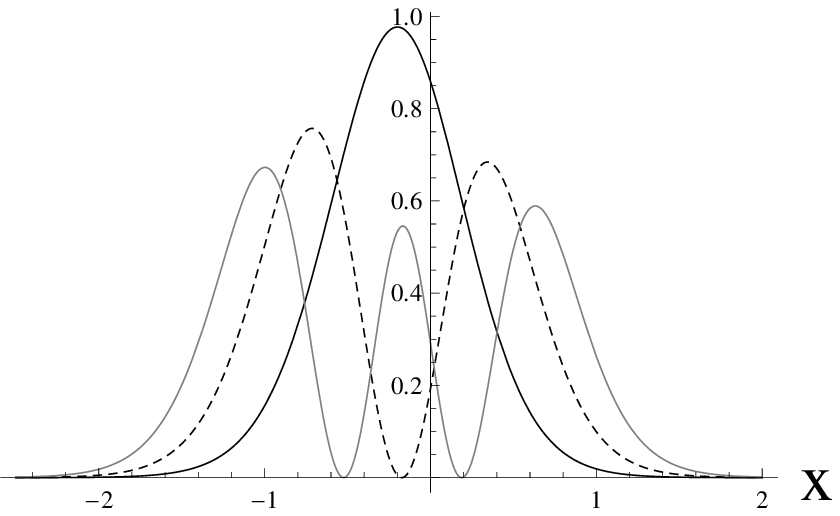,width=7.8cm}
\caption{The densities $|\psi_+|^2$ (left plot) and $|\psi_-|^2$ (right plot) for the 
values $n=0$ (black curve), $n=1$ (dashed curve), and $n=2$ (gray curve), respectively. 
Parameter settings are $\alpha=1$, $k_y=3$ and $\beta=1$.}
\label{fig2}
\end{center}
\end{figure}

We will now employ a new set of parameter values for our energy-dependent potential (\ref{v}). Even though we are using 
the same form of the potential, it will turn out that in this example the sequence of stationary energies does not increase 
linearly, but converges to zero. We make the following parameter definitions
\begin{eqnarray}
\lambda(E) ~=~ -\frac{\alpha}{E} \qquad \qquad \qquad \mu(E) ~=~ \beta, \label{set2}
\end{eqnarray}
where the $\alpha>0$ and $\beta$ are real constants. These settings comply with the condition (\ref{sign}) that 
ensures the integrand in the norm (\ref{normx}) to be a nonnegative function. Let us add that $\beta$ can be zero, such 
that this example includes the case of a hyperbolic secant potential. Next, upon substitution of the parameters 
(\ref{set2}) into our potential (\ref{v}) we obtain
\begin{eqnarray}
V(x,E) &=& \frac{\alpha}{E}~\mbox{sech}(x)+\beta~\tanh(x)+E, \label{v2}
\end{eqnarray}
In contrast to its counterpart (\ref{v1}) from the previous example, this potential has inverse energy dependence in its 
first term. We will now construct the stationary energies supported by the Dirac equation (\ref{dirac}) with potential 
(\ref{v2}). To this end, we must solve equation (\ref{eigen}) with respect to $E$, where we again choose the parameter 
value $\epsilon=1$. After incorporation of the settings (\ref{set2}) we get the condition
\begin{eqnarray}
\frac{1}{2}+\frac{\alpha}{E}+\sqrt{k_y^2-\beta^2} &=& -n,~~~|k_y|~\geq~\beta. \nonumber
\end{eqnarray}
We now obtain our stationary energies by solving for $E$. Upon renaming $E=E_n$ we arrive at 
\begin{eqnarray}
E_n &=& -\frac{2~\alpha}{2~n+1+2~\sqrt{k_y^2-\beta^2}},~~~|k_y|~\geq~\beta. \label{eigenx2}
\end{eqnarray}
These energy values are negative and increase monotically with $n$, as shown in figure \ref{fig3b}. 
Next we need to find out how many stationary 
energies are provided by (\ref{eigenx2}), let us verify that our parameter $\lambda$ satisfies the condition (\ref{l2}), 
guaranteeing existence of the 
norm (\ref{normx}). We substitute the settings (\ref{set2}) and (\ref{eigenx2}) into (\ref{l2}), arriving at
\begin{eqnarray}
\sqrt{k_y^2-\beta^2}&>&0. \nonumber
\end{eqnarray}
Since we are assuming that $|k_y|~\geq~\beta$, this condition is fulfilled for all values of our parameters. Therefore 
we have an infinite numbers of stationary energies (\ref{eigenx2}) that accumulate at $E=0$. In addition, these 
energies belong to solutions that are normalizable in the sense of our norm (\ref{normx}). These solutions can 
be constructed from the function $\psi_1$ that is defined in (\ref{psi1bound}). Upon substitution of our 
current parameter setting (\ref{set2}) and the stationary energies (\ref{eigenx2}) we obtain its explicit form
\begin{eqnarray}
\psi_{1,n}(x) &=& 
\left[1-i~\sinh(x)\right]^{-\frac{1}{4}\left[1+2n+2\sqrt{k_y^2-\beta^2}-2i\beta\right]}
\left[1+i~\sinh(x)\right]^{\frac{1}{4}\left[1+2n-2\sqrt{k_y^2-\beta^2}+2 i \beta\right]} \times \nonumber \\[1ex]
&& \hspace{-1.5cm} \times~ 
P_n^{\left(-\frac{1}{2}\sqrt{\left[2+2n+2\sqrt{k_y^2-\beta^2}+2i\beta\right]^2},-\frac{1}{2}\sqrt{\left[
-2n-2\sqrt{k_y^2-\beta^2}+2i\beta\right]^2}
\right)}\left[i \sinh(x) \right]. \label{psix2}
\end{eqnarray}
where we included an index $n$ to emphasize the bound-state character. After calculating the function $\psi_{2,n}$ 
from (\ref{psix2}), we can determine the solution spinor 
(\ref{psinew}) of our Dirac equation (\ref{dirac}) by calculating the functions $\psi_+$ and $\psi_-$. Since the 
explicit expressions of these functions are too long to be shown here, we restrict ourselves to present their graphs 
for a particular parameter setting, see figure \ref{fig2}. Recall that we do not need to verify normalizability 
according to (\ref{normx}), as this is guaranteed by (\ref{l2}) and the fact that $\mu$ is independent of the energy. 

\section{Concluding remarks}
In this work we have demonstrated how to construct bound-state solutions of the massless Dirac equation 
for an energy-dependent potential. Our approach of decoupling the Dirac equation relies on the potential depending 
on only one of the spatial variables. If this condition is fulfilled, bound states for energy-dependent 
potentials different from (\ref{v}) can be constructed, provided the Schr\"odinger-type equation (\ref{eq1}) renders 
exactly-solvable. In addition, the sign condition (\ref{sign}) for the modified norm must be verified. In most cases, this 
condition will either give restrictions on the parameters of the potential or dictate that the problem's domain must 
be restricted.

\end{sloppypar}

\end{document}